\documentclass[11pt]{article}

\usepackage{hyperref}

\input{psfig.sty}
\usepackage{fullpage,epsfig,picinpar,latexsym,amsmath,amsfonts}
\usepackage{algorithmic}
\usepackage{algorithm} 

\newcommand{\etal}{\emph{et al.}}

\newcommand{\baru}{{\bar u}}
\newcommand{\barx}{{\bar x}}

\newcommand{\barF}{{\bar F}}

\newcommand{\calB}{{\cal B}}

\newcommand{\calC}{{\cal C}}

\newcommand{\calF}{{\cal F}}

\newcommand{\calK}{{\cal K}}
\newcommand{\calL}{{\cal L}}

\newcommand{\calU}{{\cal U}}

\newcommand{\eps}{{\epsilon}}

\newcommand{\braced}[1]{{ \left\{ #1 \right\} }}

\newcommand{\parend}[1]{{ \left( #1 \right) }}

\newcommand{\cost}{\mbox{\it cost}}
\newcommand{\opt}{\mbox{\it opt}}
\newcommand{\fopt}{\overline{\mbox{\it opt}}}

\newcommand{\randcust}[1]{{r({#1})}}

\newcommand{\weight}{{w}}
\newcommand{\barweight}{{{\bar w}}}

\newcommand{\integers}{{\mathbb{Z}}}

\newcommand{\posreals}{{\mathbb{R}^+}}

\newcommand{\posnaturals}{{\mathbb{N}^+}}

\newtheorem{theorem}{Theorem}

\newtheorem{lemma}{Lemma}
\newtheorem{fact}{Fact}

\newenvironment{proof}%
   {\smallskip\noindent{\it Proof:\/} }{\hfill$\Box$\medskip}
\newenvironment{proofof}[1]%
   {\smallskip\noindent{\it Proof of {#1}:\/} }{\hfill$\Box$\medskip}
                {$\spadesuit$\smallskip}

\newenvironment{bigeqn*}{\large\begin{eqnarray*}}{\end{eqnarray*}}

\begin{document}

\title{Incremental Medians via Online Bidding\thanks{%
    The conference version of this paper appeared in \cite{ChKeNY06}. }}

\author{
       Marek Chrobak\thanks{%
       Department of Computer Science,
       University of California,
       Riverside, CA 92521.
       Research supported by NSF Grant CCR-0208856.
       }
       \and
       Claire Kenyon\thanks{%
       Computer Science Department,
       Brown University,
       Providence, RI 02912.
       }
       \and
       John Noga\thanks{%
       Department of Computer Science,
       California State University,
       Northridge, CA 91330.
       }
       \and
       Neal E. Young\thanks{%
       Department of Computer Science,
       University of California,
       Riverside, CA 92521.
       }
   }

   \date{
     \sf To appear in
     Algorithmica 50 (4), 455-–478, 2008.
     \url{https://doi.org/10.1007/s00453-007-9005-x}
   }


\maketitle

\begin{abstract}
In the $k$-median problem we are given sets of facilities and
customers, and distances between them.
For a given set $F$ of facilities, the cost of serving a
customer $u$ is the minimum distance between $u$ and a facility
in $F$. The goal is to find a set $F$ of $k$ facilities that
minimizes the sum, over all customers, of their service costs.

Following the work of Mettu and Plaxton, we study the
\emph{incremental} medians problem, where $k$ is not known in advance.
An incremental algorithm
produces a nested sequence of facility sets
$F_1\subseteq F_2\subseteq ...\subseteq F_n$,
where $|F_k| = k$ for each $k$. Such an
algorithm is called \emph{$c$-cost-competitive}
if the cost of each $F_k$ is at most $c$ times the
optimum $k$-median cost.
We give improved incremental algorithms for the metric version of 
this problem: an $8$-cost-competitive
deterministic algorithm, a $2e\approx 5.44$-cost-competitive
randomized algorithm, 
a $(24+\epsilon)$-cost-competitive, polynomial-time deterministic algorithm,
and a $6e+\eps\approx16.31$-cost-competitive, polynomial-time randomized algorithm.

We also consider the competitive ratio with respect to \emph{size}.
An algorithm is \emph{$s$-size-competitive} if the cost of each $F_k$
is at most the minimum cost of any set of $k$ facilities, while the
size of $F_k$ is at most $s k$.  We show that the optimal
size-competitive ratios for this problem, in the deterministic and
randomized cases, are $4$ and $e$.  For polynomial-time algorithms, we
present the first polynomial-time $O(\log m)$-size-approximation
algorithm for the offline problem, as well as a polynomial-time
$O(\log m)$-size-competitive algorithm for the incremental problem.
  
Our upper bound proofs reduce the incremental medians problem
to the following \emph{online bidding} problem:
faced with some unknown threshold
$T\in\posreals$, an algorithm must submit ``bids'' $b\in\posreals$
until it submits a bid $b\ge T$, paying the sum
of all its bids.  We present folklore algorithms for online bidding
and prove that they are optimally competitive.

We extend some of the above results for incremental medians to
approximately metric distance functions and to incremental fractional
medians. Finally, we consider 
a restricted version of the incremental
medians problem where $k$ is restricted to one of two given values,
for which we give a deterministic algorithm with a nearly
optimal cost-competitive ratio.
\end{abstract}


\newpage
\section{Introduction and Summary of Results}


\paragraph{The $k$-median problem.}
An instance of the $k$-median problem is specified by a finite set
$\calC$ of customers, a finite set $\calF$ of facilities, and, for
each customer $u$ and  facility $f$, a distance $d_{uf}\ge 0$ from
$u$ to $f$ representing the cost of serving $u$ from $f$.
The cost of a set of facilities $F\subseteq\calF$ is 
$\cost(F)=\sum_{u\in\calC}d_{uF}$, where
$d_{uF}=\min_{f\in{}F}d_{uf}$.  For a given $k$, 
the \emph{offline} $k$-median problem is to
compute a \emph{$k$-median}, that is,
a set $F\subseteq\calF$ of cardinality $k$ for which 
$\cost(F)$ is minimum (among all sets of cardinality $k$).
This minimum cost is denoted  $\opt_k$.
An instance of the $k$-median problem is called \emph{metric} if
 the distance function is a metric (the shortest $u$-to-$f$ path
has length $d_{uf}$ for each $u\in \calC$ and $f\in\calF$).

The $k$-median problem is a well-known NP-hard facility location
problem.  Substantial work has been done on efficient approximation
algorithms that, given $k$, find a set
$F_k$ of $k$ facilities of approximately minimum cost
\cite{arya01,arya04,ArRaSh03,charikar99a,charikar99b,charikar-guha05,%
jain01,jain02,jain03,young00kmedians,lin92packing}.
In particular, for the metric version, Arya~\etal~\cite{arya01,arya04}
show that, for any $\epsilon > 0$,
a set $F_k$ of cost at most
$(3+\epsilon)\opt_k$ can be found in polynomial time.


\paragraph{Incremental medians.}
The \emph{incremental} medians problem
is a version of the $k$-median problem
where $k$ is not specified in advance \cite{MetPla00,mettu03}.
Instead, authorizations for additional facilities arrive over time.
Given an instance of the $k$-median problem,
a (possibly randomized) algorithm produces a sequence
$\barF=(F_1,F_2,\ldots,F_n)$ of facility sets, where
$F_1\subseteq{}F_2\subseteq\cdots\subseteq{}F_n{}\subseteq\calF$,
and $|F_k|\le k$ for all $k$.

In general, in an incremental solution, the $F_k$'s cannot all
simultaneously have minimum cost. An algorithm is said to be
\emph{$c$-cost-competitive}, or to have
\emph{cost-competitive ratio} of $c$,
if it produces a (possibly random) sequence $\barF$ of sets which
is $c$-cost-competitive, that is, such that for each $k$, the set
$F_k$ has size at most $k$ and (expected) cost at most $c\cdot\opt_k$.

Alternatively, an algorithm is $s$-\emph{size}-competitive if it produces
a (possibly random) sequence $\barF$ of sets which is $s$-size-competitive,
that is, such that each set $F_k$ has cost at most $\opt_k$ and
(expected) \emph{size} at most $sk$.

For offline solutions we use the term ``approximate'' instead of
``competitive''.

\paragraph{Online bidding.}
To analyze incremental medians, we reduce the various
incremental medians problems to the following folklore 
\emph{online bidding} problem.
An algorithm repeatedly submits ``bids'' $b\in\posreals$, until it
submits a bid $b$ that is at least as large as some unknown threshold
$T\in\posreals$.  The algorithm's cost is the sum of the submitted bids.  The
algorithm is $\beta$-competitive if, for any $T\in\posreals$, its cost
is at most $\beta T$ (or, if the algorithm is randomized, its expected
cost is at most $\beta T$).  More generally, the algorithm may be
given in advance a closed universe $\calU\subseteq\posreals$, with a
guarantee that the threshold $T$ is in $\calU$ and a requirement
that all bids be in $\calU$. 
(To handle the case when $\calU$ includes 
arbitrarily small positive numbers, we allow the bidding sequence
to be left-infinite -- see Section~\ref{sec: online bidding}
for a formal definition.)

In Section~\ref{sec: online bidding} we completely characterize
optimal competitive ratios for deterministic and randomized algorithms
for online bidding:


\begin{theorem}\label{thm: online bidding upper}
{\rm\bf [folklore]}
{\rm\bf (a)}
The online bidding problem has a deterministic $4$-competitive
algorithm. If the universe $\calU$ is finite, this algorithm
runs in time polynomial in $|\calU|$.
{\rm\bf (b)}
The online bidding problem has a randomized $e$-competitive
algorithm. If $\calU$ is finite, this algorithm
runs in time polynomial in $|\calU|$.
\end{theorem}

Throughout, $[n]$ denotes $\{1,2,\ldots,n\}$.
\begin{theorem}\label{thm: online bidding lower}
{\rm\bf (a) [folklore]}
No deterministic algorithm for online bidding is less than $4$-competitive,
even when restricted to
instances of the form $\calU=[n]$ for some integer $n$.
{\rm\bf (b)}
No randomized algorithm for online bidding is less than $e$-competitive,
even when restricted to
instances of the form $\calU=[n]$ for some integer $n$.
\end{theorem}

Portions of Theorems~\ref{thm: online bidding upper}
and~\ref{thm: online bidding lower} are folklore. In particular,
Theorem~\ref{thm: online bidding upper}(a) uses a doubling
algorithm that has been used previously in several papers,
first in~\cite{krt96,motwani94nonclairvoyant} and later 
in~\cite{gk96,goemans-kleinberg98,chakrabarti96improved,cfm97,dl05}. 
Some of these papers also have the randomized upper bound.
We include proofs of all bounds for completeness.
Our main new contribution in this setting is 
Theorem~\ref{thm: online bidding lower}(b),
a randomized lower bound that matches the known upper bound. 


\paragraph{Size-competitiveness.}
To our knowledge, size-competitive algorithms for incremental medians
have not been studied, although other online problems have been
analyzed in an analogous setting of
\emph{resource augmentation} (e.g. \cite{KalPru00,ChGoKK04,Koutso99}).
For unrestricted (possibly non-polynomial-time) algorithms, 
we completely characterize the optimal size-competitive ratios:

\smallskip
\begin{theorem}\label{thm: size competitive deterministic}
{\rm\bf (a)}
The incremental medians problem has
a $4$-size-competitive deterministic algorithm.
{\rm\bf (b)}
No deterministic incremental algorithm has 
size-competitive ratio smaller than $4$.
\end{theorem}


\begin{theorem}\label{thm: size competitive randomized}
{\rm\bf (a)}
The incremental medians problem has
an $e$-size-competitive randomized algorithm.
{\rm\bf (b)}
No randomized incremental algorithm has 
size-competitive ratio smaller than $e$.
\end{theorem}

We stress that the upper and lower bounds in
Theorems~\ref{thm: size competitive deterministic}
and~\ref{thm: size competitive randomized} 
are for unrestricted algorithms
and hold for both the metric and non-metric problems.

Regarding polynomial-time algorithms, previously
no polynomial-time size-approximation algorithms 
for $k$-medians were known, even for the offline
problem.  The best previous result for the offline
problem is a bicriteria-approximation algorithm
which finds a facility set of size $O(k\log(m+m/\eps))$ 
and cost at most $(1+\eps)\opt_k$ \cite{young00kmedians} 
(improving on the previous work in 
\cite{lin92,lin92packing,korupolu00,young00kmedians}).
We first improve this result to obtain an offline size-approximation algorithm:


\begin{theorem}\label{thm: size competitive offline}
The offline $k$-median problem has a polynomial-time
$O(\log m)$-size-approximation algorithm,
where $m=|\calC|$ is the number of customers.
\end{theorem}

Note that this algorithm finds a true (not bicriteria) approximate solution:
a facility set of size $O(k \log m)$ and cost at most $\opt_k$.
We use this result and a reduction to give a polynomial-time
size-approximation algorithm for the incremental problem:


\begin{theorem}\label{thm: size competitive polynomial}
  The incremental medians problem has a polynomial-time
  $O(\log m)$-size-competitive algorithm.
\end{theorem}

The bounds in Theorems~\ref{thm: size competitive offline} 
and~\ref{thm: size competitive polynomial} are optimal in the sense that
no polynomial-time algorithm (incremental or offline) is
$o(\log m)$-size-competitive unless P=NP, even for the metric case.
This follows from known results on the hardness of approximating set cover.

Theorems~\ref{thm: size competitive deterministic},
\ref{thm: size competitive randomized},
and~\ref{thm: size competitive polynomial}
(proven in Section~\ref{sec: size-competitive medians})
imply the size-competitive ratios shown in Figure~\ref{fig: results}.

\begin{figure}[t]\small
  \centering
\begin{tabular}[h]{|r|c|c|c|c|c|}\hline
\em problem:
&
\multicolumn{2}{|c|}{\bf cost-competitive, metric} 
&
\multicolumn{2}{|c|}{\bf size-competitive}
&{\bf bidding}
\\  \hline
\em time:
&{polynomial} 
&{unrestricted}
&{polynomial} 
&{unrestricted}
&polynomial
\\\hline
deterministic     & $24+\eps$     &  $8$ 
& ${\mbox{\boldmath $O(\log m)$}}$
& {\bf 4}
& {\bf 4}
\\\hline
randomized        & $6e+\eps< 16.31$  &  $2e< 5.44$ 
& ${\mbox{\boldmath $O(\log m)$}}$
& ${\mbox{\boldmath $e$}} < 2.72$
& ${\mbox{\boldmath $e$}} < 2.72$
\\ \hline
\end{tabular}
  \caption{\small Competitive ratios shown for incremental medians and online bidding.  Ratios in bold are optimal.}
  \label{fig: results}
\vspace*{-0.1in}
\end{figure}


\paragraph{Cost-competitive incremental medians.}
For incremental medians, 
Mettu and Plaxton~\cite{MetPla00,mettu03} give a $c$-cost-competitive
linear-time algorithm with $c\approx 30$.  We improve this result.
The problem is difficult
both because (i) the solution must be
incremental, and (ii) even the offline problem is NP-hard.    
To study separately the effects of the two difficulties,
we consider both polynomial-time and unrestricted
algorithms, proving the following two theorems
(see Section~\ref{sec: cost-competitive medians} for the proofs):


\smallskip

\begin{theorem}\label{thm: cost competitive deterministic}
{\rm\bf (a)} The metric incremental medians problem has
an $8$-cost-competitive deterministic algorithm.
{\rm\bf (b)} Suppose that the offline metric $k$-median problem
has a polynomial-time $c$-cost-approximation algorithm. Then the
incremental medians problem has a polynomial-time 
$8c$-cost-competitive deterministic algorithm.
\end{theorem}


\begin{theorem}\label{thm: cost competitive randomized}
{\rm\bf (a)} The metric incremental medians problem has
a $2e$-cost-competitive randomized algorithm.
{\rm\bf (b)} Suppose that the offline metric $k$-median problem
has a polynomial-time $c$-cost-approximation algorithm. Then the
incremental medians problem has a polynomial-time 
$2ec$-cost-competitive randomized algorithm.
\end{theorem}

As it is known that there is a polynomial-time
$(3+\eps)$-cost-approximation algorithm for the metric
$k$-median problem \cite{arya01,arya04},
Theorems~\ref{thm: cost competitive deterministic}
and~\ref{thm: cost competitive randomized} imply the
cost-competitive ratios shown in Figure~\ref{fig: results}.

Theorems~\ref{thm: cost competitive deterministic} 
and~\ref{thm: cost competitive randomized} were recently
and independently discovered by
Lin, Nagarajan, Rajaraman and Williamson~\cite{LNRW06}.
For polynomial-time algorithms, 
they improve the cost-competitive ratios further
to $16$ and $4e$, in the deterministic and randomized
cases, respectively.
(The general approach in \cite{LNRW06} is based on the
doubling method similar to ours; the improvements were
accomplished using a Lagrangian-multiplier-preserving 
approximation for facility location.)
They also  generalize the approach to incremental
versions of $k$-MST, $k$-vertex cover,  $k$-set cover, facility location,  
and to hierarchical $k$-median.


\paragraph{$\lambda$-Relaxed metrics.}
Mettu and Plaxton show that their incremental medians algorithm also works
in ``$\lambda$-approximate'' metric spaces, achieving
cost-competitive-ratio $O(\lambda^4)$ \cite{MetPla00,mettu03}.
We get a similar result. Let $\lambda\ge 1$.
We say that the cost function
$d$ is a \emph{$\lambda$-relaxed metric} if $d_{fy} \le \lambda
(d_{fx}+ d_{xg}+d_{gy})$ for any facilities $f,g$ and customers $x$ and $y$.
(This condition is somewhat less restrictive than the one
in \cite{MetPla00,mettu03}. A related concept was studied in
\cite{faginstockmeyer98}.)
In Section~\ref{sec: lambda-relaxed metrics}, we prove that
Theorems~\ref{thm: cost competitive deterministic}
and~\ref{thm: cost competitive randomized}
generalize as follows:


\smallskip

\begin{theorem}\label{thm: cost competitive lambda deterministic}
  {\rm\bf (a)} The incremental medians problem
  for $\lambda$-relaxed metrics has a
  deterministic algorithm with cost-competitive ratio
  $8\lambda^2$.
  {\rm\bf (b)} If the offline $k$-median problem for
  $\lambda$-relaxed metrics has a
  polynomial-time $c$-cost-approximation algorithm, then the incremental
  version has a deterministic polynomial-time algorithm
  with cost-competitive ratio $8\lambda^2c$.
\end{theorem}

\begin{theorem}\label{thm: cost competitive lambda randomized}
  {\rm\bf (a)} The incremental medians problem
  for $\lambda$-relaxed metrics has
  a randomized algorithm with cost-competitive ratio
  $2e\lambda^2$.
  {\rm\bf (b)} If the offline $k$-median problem for
  $\lambda$-relaxed metrics has a
  polynomial-time $c$-cost-approximation algorithm, then the incremental
  version has a randomized polynomial-time algorithm
 with cost-competitive ratio
  $2e\lambda^2c$.
\end{theorem}


\paragraph{Incremental fractional medians.}
A \emph{fractional $k$-median} $x$ is a solution to the standard
linear program relaxation for the $k$-median problem
(see Section~\ref{sec: fractional medians}). In this linear program, 
$x_{uf}$ specifies how much of the demand from customer $u$
is served by facility $f$; thus we have a
constraint $\sum_{f\in\calF} x_{uf} = 1$.
For each $f\in \calF$, the \emph{capacity required at $f$}
is $|x|_f = \max_{u\in\calC} x_{uf}$,
and the \emph{total capacity} of $x$ is
$|x| = \sum_{f\in\calF} |x|_f$.
(Naturally, this corresponds to the cardinality of $x$
in the integral case.) We require that $|x|\le k$, and
the objective is to minimize the cost of $x$, defined by
$\cost(x) = \sum_{u\in\calC,f\in\calF} d_{uf}x_{uf}$.

For two fractional medians $x$, $x'$, we say that $x$
\emph{dominates} $x'$ if $|x'|_f\le |x|_f$ for each facility $f\in\calF$.
An \emph{incremental fractional median} is defined
by a sequence $(x^{k})_k$ of fractional $k$-medians,
one for each $k\in[n]$, where each $x^{k+1}$ dominates $x^k$, for $k>n$.
This sequence is $c$-cost-competitive if
$\cost(x^k)\le c\cdot \fopt_k$ for each $k$,
where $\fopt_k$ is the minimum cost of
of any (non-incremental) fractional $k$-median.

To prove the theorem below (see Section~\ref{sec: fractional medians}),
we extend the proof of
Theorem~\ref{thm: cost competitive randomized} to the fractional case,
then note that the randomized algorithm for the
fractional problem can be derandomized without increasing the
competitive ratio. 


\smallskip

\begin{theorem}\label{thm: fractional incremental median}
If the distance function is metric then there is a
deterministic polynomial-time algorithm that
produces a $2e$-cost-competitive incremental fractional median.
\end{theorem}

One motivation for introducing fractional incremental medians
(in addition to its own independent interest)
is the possibility that the above theorem, or its
improvements, could be used to improve the
$c$-cost-competitive ratio for the integral case.
Note that $\fopt_k\le\opt_k$, so if we could somehow
round a fractional $c$-competitive incremental median
to an integral solution, giving up a factor of, say,
$3$ in the cost, then we would have a $3c$-cost-competitive
deterministic algorithm for incremental medians.
(See \cite{bucNao06} for a similar approach
for online problems.)
However, the ratio $2e$ above is insufficient for this,
in that even with $c=2e$, the resulting ratio $3c=6e$ is
larger than the current best ratio of $16$.
Note that we are unlikely to lose less than a factor
of $3$ in rounding the fractional incremental median,
as it is at least
as hard as rounding a fractional $k$-median.


\paragraph{The $kl$-medians problem.}
A natural question to ask is whether better competitive
ratios are possible if the number of facilities can take
only some limited number of values. As shown in \cite{MetPla00,mettu03},
no algorithm can be better than $2$-competitive even when
there are only two possible numbers of facilities, either $1$
or $k$, for some large $k$. 

For any $1\le k < l\le n$,
we define the \emph{$kl$-medians} problem as the
restricted version of the incremental medians problem
where the number of facilities is either $k$ or $l$.
In Section~\ref{sec: kl-medians}, 
we determine almost exactly the competitive ratio of the
$kl$-medians problem in the deterministic case:

\smallskip

\begin{theorem}\label{thm: kl-medians deterministic}
Let $1\le k < l \le n$.
{\rm (a)} If the distance function is metric, then
there is a deterministic $kl$-median algorithm 
with cost-competitive ratio $2-1/l$, and
{\rm (b)}
no ratio better than $2-1/(l-k+1)$ is possible.
\end{theorem}


\paragraph{Bicriteria approximations.}
We say that an algorithm for the $k$-median problem is a
\emph{bicriteria $(c,s)$-approximation}
algorithm if it approximates the cost within the
ratio of $c$ and the size within the ratio of $s$.
Such bicriteria approximation algorithms for $k$-medians
have been studied by many authors, 
\cite{lin92,lin92packing,korupolu00,young00kmedians}.
We remark without proof that, analogously to
Theorem~\ref{thm: cost competitive deterministic} and
Theorem~\ref{thm: size competitive deterministic},
one can transform any offline polynomial-time bicriteria
$(c,s)$-approximation
algorithm into a polynomial-time \emph{incremental} algorithm
whose bicriteria performance guarantee is either
$(c,4s)$ or $(8c,s)$. For example, for metric spaces,
using the approximation results from
\cite{arya01,arya04,lin92,korupolu00} one can obtain incremental
polynomial-time algorithms 
with the following respective bicriteria ratios:
$(3+\epsilon,4)$,
$(2+\epsilon , 4(1+2\epsilon^{-1}))$, and
$(1+\epsilon , 4(3+5\epsilon^{-1}))$, where $\epsilon > 0$.


\paragraph{Weighted medians.}
It is quite easy to see that
all of the results in this paper extend to the version of $k$-medians
where customers are given non-negative weights.
Denote by $w_u$ the weight of $u \in \calC$.
In this generalization, the cost of a facility set
$F$ is defined by $\cost(F) = \sum_{u\in\calC}w_ud_{uF}$.
The results on size-competitive algorithms can be
further extended to the case where each facility $f$
is assigned a weight $w_f$. The definition of the
cost function remains the same. The value $k$ represents
now an upper bound on the allowed total facility weight 
$\sum_{f\in F} w_f$.


\section{Online Bidding}
\label{sec: online bidding}

In this section we provide a complete analysis of
online bidding by proving Theorems~\ref{thm: online bidding upper} 
and~\ref{thm: online bidding lower}.

Throughout the paper, $\posreals$ denotes the set of non-negative
reals, $\integers$ the set of integers, and $\posnaturals$ the set of
positive integers.  For $n\in\posnaturals$, let
$[n]=\braced{1,2,\dots,n}$.

Given a universe $\calU$ which is a closed subset of $\posreals$,
an {\em online bidding} algorithm outputs a \emph{bid set}
$\calB\subseteq\calU$. Against a particular threshold $T\in\calU$, the
algorithm's cost is
\begin{eqnarray*}
\sum \{b\in\calB : b < T\}
        + 
        \min \{b\in\calB : b \ge T\}.
\end{eqnarray*}
The bid set $\calB$ is
\emph{$\beta$-competitive} if, for any $T\in\calU$,
this cost is at most $\beta T$.


\smallskip

\begin{proofof}{Theorem~\ref{thm: online bidding upper}(a)}
We give a deterministic $4$-competitive algorithm.

First, consider the case $\calU=\posreals$.  Define the
algorithm to produce the set of bids
$\calB = \{0\}\cup\{2^i:i\in\integers\}$.
If the threshold $T$ is zero, the algorithm pays zero.
For any other threshold $T>0$, let $p$ be such that
$T\in (2^{p-1}, 2^p]$. 
The algorithm pays $\sum_{i \le p} 2^i = 2^{p+1} \le 4T$,
and thus its competitive ratio is at most $4$.
  
Next, we reduce the general case  to the case
$\calU=\posreals$.   Let $\calB$ be a 4-competitive
bid set for $\calU=\posreals$.  For an arbitrary closed
 universe $\calU'\subseteq\posreals$,
the algorithm produces the bid set
\begin{equation*}
\calB' \;=\; \braced{ \max(\calU'\cap[0,b]) \,:\, b\in\calB}.
\end{equation*}
In other words, we replace each 
$b\in\calB$ with the maximum element in $\calU'\cap [0,b]$ (if any).   
The cost incurred against any threshold $T\in\calU'$
is at most the cost incurred when using the bid set $\calB$
against the same threshold $T$.
Thus, the bid set $\calB'$ is also $4$-competitive.

Note that if $\calU$ is finite then for the bid set $\calB$ 
described in the previous paragraph, the corresponding
bid set $\calB'$ can be 
computed in time polynomial in $|\calU|$.
\end{proofof}


\begin{proofof}{Theorem~\ref{thm: online bidding upper}(b)}
We give a randomized $e$-competitive algorithm. 

First, we consider the case $\calU=\posreals$.  
The algorithm picks a real number $\xi\in [0,1)$
uniformly at random, then produces the set of bids
$\calB=\{0\}\cup\{e^\xi e^{i}:i\in\integers\}$.

We now give the analysis of this algorithm.
We first observe that for any integer
$x$, the set $\calL = \{\ln(b) \,:\, b\in\calB-\braced{0}\}$
induces a uniform distribution in the interval $[x,x+1)$,
in the sense that $[x,x+1)$ contains exactly one element of $\calL$
and this element is uniformly distributed in $[x,x+1)$.
This immediately implies that the above property holds
in fact for any real $x$.
 
Let $T>0$ be the threshold.
For any constant $\tau$, by the paragraph above, we can equivalently
describe the algorithm as producing the set of bids
$\calB=\{0\}\cup\{e^{\xi+ \tau + i}:i\in\integers\}$  
where $\xi$ is distributed uniformly in $[0,1)$.
In particular, we can take $\tau = \ln(T)$.

Let $b$ be the random variable equal to the largest bid
paid by the algorithm, defined by $e^{\xi+ \ln (T) + i-1}< T\leq
e^{\xi+ \ln (T) + i}=b$, or, in other words, 
$e^{\xi +i-1}<e^0\leq e^{\xi +i}=b/T$. Thus $i=0$ and
  $b/T$ is distributed like $e^{\xi}$
  with $\xi$  uniform in $[0,1)$. It follows that
  the expectation of $b$ is $T\int_0^1 e^z \,dz = T(e-1)$, 
thus the expected total payment incurred by the algorithm is  
  $E[\sum_{i=0}^\infty b e^{-i}] = E[b]e/(e-1)=eT$,
  and the algorithm is $e$-competitive.
  
The general case, for an arbitrary closed universe, reduces to
the case $\calU=\posreals$ just as in the proof of
Theorem~\ref{thm: online bidding upper}(a) above.
\end{proofof}


\begin{proofof}{Theorem~\ref{thm: online bidding lower}(a)}
We now show a deterministic lower bound matching the upper bound.

In the simple case where $\calU=\posnaturals$,
assume (towards a contradiction) that there exists an online
algorithm with competitive ratio $a < 4$. Let 
$\calB = \braced{b_i}_{i\geq 1}$ be the bid set produced by the algorithm,
$s_j = \sum_{i=1}^j b_i$ and
$y_j = s_ {j+1}/s_j$, for $j\geq 1$.
Against the threshold $T=b_j+1$,  the
algorithm pays $s_{j+1}$. By our assumption, 
$s_{j+1} \le a (b_j +1) = a(s_j-s_{j-1} +1)$  for all $j$.
Rearranging, and using $s_j\geq j$, we get:
\begin{eqnarray}
(y_{j+1}-y_j)y_j
		&\le& -y_j^2+a(s_{j+1}-s_j+1)y_j/s_{j+1}
			\nonumber	\\
         &=& -y_j^2+ay_j-a(1-1/s_j)
			\nonumber	\\
        &\leq& -y_j^2+ay_j-a(1-1/j).
			\label{eqn: bound on y}
\end{eqnarray}
For $a<4$ and $j$ large enough, the discriminant $a^2-4a(1-1/j)$ is
negative, thus the last expression is negative, and so the
sequence $(y_{j})$ is ultimately 
decreasing. As it is bounded below by 1, it converges to some limit $y$
which, by continuity, must satisfy
$0=(y-y)y\leq -y^2+ay-a$ -- a contradiction, since
$-y^2+ay-a < 0$ for $0 < a < 4$ and any $y$.

In the case when $\calU=[n]$, for arbitrarily large $n$,
the proof  is similar but a bit more explicit.
Let $\calB = \braced{b_i}_{i=1}^m$. We start out the same.
Note that $s_{j+1} \le a (b_j +1)$ implies $s_{j+1}<
8 s_j$, and so $y_j< 8$ for all $j = 1,...,m-1$.

Let $j_0$ be the smallest positive integer such that
$a^2-4a(1-1/j_0)=-\delta<0$. 
Inequality~(\ref{eqn: bound on y}) is valid for $j = j_0,...,m-2$, 
so we have:
\begin{equation*}
(y_{j+1}-y_j)y_j
        \;\leq\; \max_y (-y^2+ay-a(1-1/j_0))
        \;=\; -{\delta / 4}.
\end{equation*}
Since $y_j\leq 8$, the above inequality 
implies that $y_{j+1}-y_j\leq -\delta/32$.
Therefore, for $j = j_0,...,m-1$, we have
$1\le y_j \le y_{j_0} - (j-j_0)\delta/32 \le 8 - (j-j_0)\delta/32$,
so, in particular, $m \le j_0 + 1 + 224/\delta$.
Since $b_m = n$ (otherwise, $\calB$ would not be
competitive at all), $b_1\le 8$, and $b_{j+1}\le 8b_j$ for all $j=1,...,m-1$,
we get $n \le 8^{j_0+1+224/\delta}$,
contradicting our assumption that $n$ can be arbitrarily large.
\end{proofof}


\begin{proofof}{of Theorem~\ref{thm: online bidding lower}(b)}
We now show a randomized lower bound matching the upper bound.
The proof consists of the two lemmas below.

\begin{lemma}\label{duality}
  Let $n\in\posnaturals$. Suppose that there are
  $\mu:[n] \rightarrow \posreals$ and
  $\pi:[n] \rightarrow \posreals$ that satisfy
  \begin{equation}\label{eqn: dual}
  \sum_{T=t}^n \frac{1}{T}\pi(T) ~\ge~ 
  \frac{1}{b} \sum_{T=t}^{b} \mu(T)
  ~~~\forall b,t: 1\le t \le b \le n.
  \end{equation}
Then any randomized $\beta$-competitive online
bidding algorithm for $\calU=[n]$ must have
\begin{equation}\label{eqn: dual bound}   
\beta ~\geq~ \frac{\sum_{T=1}^n \mu(T)}{\sum_{T=1}^n \pi(T)}.
\end{equation}
\end{lemma}

\begin{proof}{}
  Consider a $\beta$-competitive
  randomized algorithm for $\calU$, and
  let $\calB=\{b_1,b_2,\ldots,b_m\}$ be the ordered (random) sequence of bids 
produced by the algorithm. Without loss of generality, $b_m=n$.

For $t\leq b$, let $X(t,b)$ be the characteristic function of the event that $t-1$ and $b$
are consecutive elements of $\{ 0\} \cup \calB$. 
The algorithm pays $\sum_{t\leq T}\sum_b bX(t,b)$ against threshold $T$.
Since it is $\beta$-competitive, this is at most $\beta T$ in expectation.
Since the algorithm only stops when reaching a bid greater than or equal to $T$,
we must always have $\sum_{t,b:t\leq T\leq b}X(t,b)\geq 1$. Hence, together
with $\beta$, the expectation
$x(t,b)$ of $X(t,b)$ must form a feasible solution
 to the following linear program (LP):
  \[\mbox{minimize}_{\beta,x}~ \beta \mbox{~~subject to~}
  \left\{\begin{array}{rcll}
      \displaystyle
      \beta - \sum_{b=1}^n \frac{b}{T} \sum_{t=1}^T x(t,b) & \ge & 0
    & \forall T\in[n]
    \\
    \displaystyle
    \sum_{b=T}^n\sum_{t=1}^T x(t,b) & \ge & 1 
    & \forall T\in[n]
    \\
    x(t,b) & \ge & 0  
	&\forall t,b\in[n].
  \end{array}\right.
\]

  Thus, the value of  (LP) is a lower bound on the
  optimal competitive ratio of the randomized algorithm.
  To get a lower bound on the value of (LP), we use the dual (DLP)
(where the dual variables $\pi(T)$ correspond to the first set of
constraints and $\mu(T)$ to the second set of constraints):
  \[
    \mbox{maximize}_{\mu,\pi}~ \sum_{T=1}^n \mu(T)~
    \mbox{~~subject to~}
    \left\{\begin{array}{rcll}
    \displaystyle
    \sum_{T=1}^n \pi(T) & \le & 1 
    \\
    \displaystyle
    \sum_{T=t}^b \mu(T)~
    - \sum_{T=t}^n \frac{b}{T}\,\pi(T)~ &\le& 0
    & \forall t,b\in[n]
    \\
    \mu(T),\pi(T)&\ge& 0 &
      \forall T\in[n].
  \end{array}
  \right.
  \]

  Now, given any $\mu$ and $\pi$ meeting the condition of the lemma,
  if we scale $\mu$ and $\pi$ by dividing by $\sum_T \pi(T)$, we
  get a feasible dual solution whose value is 
  $\sum_T \mu(T)~/ \sum_T \pi(T)$, and the lemma follows.
\end{proof}


\begin{lemma}\label{lem: bidding rand lb}
  There exist $\mu:[n]\rightarrow\posreals$ and
  $\pi:[n]\rightarrow\posreals$ satisfying
  Condition~(\ref{eqn: dual}) of Lemma~\ref{duality}
  and such that $\sum_{T=1}^n \mu(T) / \sum_{T=1}^n \pi(T) \ge (1-o(1))e$ as $n\to\infty$.
\end{lemma}

\begin{proof}{}
  Fix $U$ arbitrarily large and let $n=\lceil U^2\log U \rceil$.
  Let $\alpha>0$ be a constant to be determined later.
  We will choose $\alpha$ so that Condition~(\ref{eqn: dual}) 
  holds, and then show that the corresponding lower bound is
  $e(1-o(1))$ as $U\rightarrow\infty$.
  Define
  \[\mu(T) = \begin{cases}
    \alpha/T & \textrm{if\ } U \le T \le U^2 \cr
    0 & \textrm{otherwise}
  \end{cases}
  \mbox{~~~~and~~~}
  \pi(T) = \begin{cases}
    1/T & \textrm{if\ } U \le T \le U^2\log U \cr
    0 & \textrm{otherwise.}
  \end{cases}.\]

  If $t> U^2$,
  then the right-hand side of Condition~(\ref{eqn: dual}) has value 0, so the condition holds
  trivially.  On the other hand,  since $\pi(T)$ and $\mu(T)$
  are zero for $T<U$, if the condition holds for $t=U$, then it also holds
  for $t<U$.  So, we need only verify the condition for $t$ in the
  range $U\le t \le U^2$.
  The expression on the left-hand side of~(\ref{eqn: dual}) then has value
  \[ 
  \sum_{T=t}^{U^2\log U} \frac{1}{T^2}
  ~\ge~
  \int_{t}^{1+U^2\log U} \frac{1}{T^2}~dT
   ~=~
  \frac{1}{t} - \frac{1}{1+U^2\log U} 
  ~\ge~
  \frac{1}{t} \parend{ 1 - \frac{U^2}{1+U^2\log U} }
  ~\ge~
  \frac{1}{t}(1 - o(1)).
  \]
  In comparison, the expression on the right-hand side has value at
  most
  \[
  \max_{b\ge t}~ \frac{1}{b} \sum_{T=t}^b \frac{\alpha}{T}~
  ~\le~
  \alpha
  \max_{b\ge t}~ \frac{1}{b} \int_{t-1}^b \frac{1}{T}~dT
  ~=~
  \alpha \max_{b\ge t}~ \frac{1}{b} \ln\frac{b}{t-1}
  ~=~
  \frac{\alpha}{e\,t(1-o(1))}.
  \]
(The second equation follows by calculus, for
         the maximum occurs when $b=e(t-1)$.)
Thus, Condition ~(\ref{eqn: dual})  is met for $\alpha = (1-o(1))e$.
Denoting by $H_j$ the $j$th harmonic number, we have
$H_j = (1+o(1))\ln(j)$.
Using this and the bounds derived above we get 
\begin{eqnarray*}
  \frac{\sum_{T} \mu(T)}{\sum_{T} \pi(T)}
  	&=&
  		\frac{\sum_{T=U}^{U^2} \alpha/T}{\sum_{T=U}^{U^2\log U} 1/T}
				\\
	&=&
		(1-o(1))e \cdot \frac{ H_{U^2} - H_{U-1} }{ H_{U^2\log U} - H_{U-1} }
				\\
	&\ge&
  		(1-o(1))e \cdot \frac{\ln (U^2/(U-1)) }{\ln( (U^2\log U)/ (U-1))}
  	~\ge~
  			(1-o(1))e.
\end{eqnarray*}
completing the proof of the lemma.
\end{proof}

Using the functions $\mu$ and $\pi$ from Lemma~\ref{lem: bidding rand lb}
in Lemma~\ref{duality}, we obtain a lower bound of $\beta \ge (1-o(1))e$ on
the competitive ratio, and
Theorem~\ref{thm: online bidding lower} follows.
\end{proofof}


\section{Incremental Size-Competitive Medians}
\label{sec: size-competitive medians}

In this section we prove
Theorems~\ref{thm: size competitive deterministic},~\ref{thm: size competitive randomized},~\ref{thm: size competitive offline}, and~\ref{thm: size competitive polynomial}.
The proofs are based on the reduction
shown in the next lemma.   
We show that from a $\beta$-competitive algorithm
for online bidding, and a 
$c$-size-approximation algorithm for
the offline $k$-median problem, we can construct
a $c\beta$-size-competitive
algorithm for the incremental medians problem.
This reduction works even for the non-metric case.
The reduction takes polynomial time, so, 
if the offline size-approximation algorithm runs in polynomial time,
the reduction yields a polynomial-time size-competitive algorithm.
Likewise, if the online bidding algorithm is randomized
then the size-competitive algorithm will be randomized.


\begin{lemma}\label{lem: size reduction deterministic}
Assume that for each $k$ we have a set 
of facilities $F_k^*$ of size at most $s k$
and cost at most $\opt_k$. Let $\beta\ge 1$.

Suppose that there exists a (randomized) polynomial-time
$\beta$-competitive algorithm for online bidding.
Then, in (randomized) polynomial time,
we can compute an \emph{incremental} solution
$\barF = (F_1,...,F_n)$ where
each $F_k$ has cost at most $\opt_k$ and
(expected) size at most $\beta s k$.
\end{lemma}

\begin{proof} 
We give the proof for the deterministic case. (The proof in the randomized
setting is an easy extension, and we omit it.)
Let $\calU=[n]$ and
take $\calB$ to be the set of bids used by the $\beta$-competitive 
online bidding algorithm  for universe $\calU$.
Let $\calB_k$ be the set of bids issued
against threshold $T=k$. We define $F_k = \bigcup_{b\in \calB_k} F_b^*$. 
Note that $\calB_j \subseteq \calB_{j+1}$ for all $j<n$,
and so $\barF$ is indeed an incremental solution.
Further, $F_k$ contains $F^*_b$ for some $b\ge k$, so $\cost(F_k)
\le \cost(F^*_b) \le \opt_b \le \opt_k$.  
Finally,  $|F_k| \le
\sum_{b\in \calB_k} |F_b^*| \le \sum_{b \in \calB_k} s b \le s \beta k$
since $\calB$ is $\beta$-competitive.
\end{proof}

\begin{proofof}{Theorems~\ref{thm: size competitive deterministic}(a)
    and~\ref{thm: size competitive randomized}(a)}
By Theorems~\ref{thm: online bidding upper}(a),
and~\ref{thm: online bidding lower}(a),
there are deterministic and randomized algorithms
for online bidding with competitive ratios of $4$ and $e$,
respectively.
Using these online bidding algorithms, and 
taking each $F^*_k$ to be the optimal $k$-median,
the reduction in Lemma~\ref{lem: size reduction deterministic}
gives a $4$-size-competitive deterministic algorithm
and an $e$-size-competitive randomized algorithm
for incremental medians.
\end{proofof}

Next, we turn our attention to the proof of
Theorem~\ref{thm: size competitive deterministic}(b),
our lower bound on the
size-competitiveness of unrestricted algorithms for incremental
medians.  We do this by showing the converse of the reduction in
Lemma~\ref{lem: size reduction deterministic}.  That is, we show that
competitive online bidding reduces to size-competitive incremental
medians:


\begin{lemma}\label{lem: size lb}
Let $s\geq 1$ and assume that, for incremental medians (metric or not),
there is a (possibly randomized)
$s$-size-competitive algorithm.  Then, for any integer $n$, 
there is a (randomized) $s$-competitive algorithm
for online bidding with $\calU=[n]$.
\end{lemma}

\begin{proof}
  We give the proof in the deterministic setting.  (The proof in the randomized
  setting is an easy extension, and we omit it.)
  For any arbitrarily large $m$,  we construct a set $\calC$
of customers, a set $\calF$ of facilities, and a metric distance
function $d_{uf}$, for $u\in\calC$ and $f\in\calF$. 
The facility set $\calF$ will be partitioned 
into disjoint sets $M_1,M_2,\dots, M_m$,
where $|M_k| = k$ for each $k$, with the following property:
\begin{description}
\item{$(\ast)$}
    For all $k$, and for every set $F$ of facilities,
    if $\cost(F)\le \cost(M_k)$ then there exists
    $\ell\ge k$ such that $M_\ell \subseteq F$.
\end{description}
Notice that the condition $(\ast)$ implies that
$\cost(M_k) > \cost(M_{k+1})$ for $k<m$, and that
each $M_k$ is the unique optimum $k$-median.
  
Assume for the moment that there exists such a metric space,
and consider an $s$-size-competitive incremental median
$\barF = (F_1,...,F_m)$ for it. This means that for each $k$
we have $|F_k|\le\beta k$ and that $F_k$ satisfies condition $(\ast)$.

Let $\calB = \{k:M_k\subseteq{}F_k\}$. We show that $\calB$ is an
$s$-competitive bid set for universe $\calU=[m]$.
For $k=m$, $(\ast)$ implies that $M_m\subseteq F_m$, and thus
$\calB\neq\emptyset$. Against any
threshold $T\in[m]$, the total of the bids paid will be
\begin{eqnarray*}
C \;=\; \sum\{ k : k < T, M_k\subseteq F_k\}
        + \min\{\ell : \ell \ge T, M_\ell \subseteq F_\ell\}.
\end{eqnarray*}
Now, $\sum\{ k : k < T, M_k\subseteq F_k\}
        \le \sum\{ k : k < T, M_k\subseteq F_T\}$ since
$\barF$ is a nested sequence.  Similarly, 
we have
$\min\{\ell : \ell \ge T, M_\ell \subseteq F_\ell\}
        \;\le\; \min\{\ell : \ell \ge T, M_\ell \subseteq F_T\}$.
Note that by $(\ast)$, $M_\ell \subseteq F_T$ for some $\ell\ge T$,
so the minimum on the right is well-defined for $T\in[m]$.
Thus we can bound the cost of $\calB$ as follows:
\begin{align*}
 C &\;\leq\; \sum\{ k : k < T, M_k\subseteq F_T\}
                +\min\{\ell : \ell \ge T, M_\ell \subseteq F_T\}\hspace*{-1in}
                &
                        \\
  &\;=\; \sum\{ |M_k| : k < T, M_k\subseteq F_T\}
                +\min\{|M_\ell| : \ell \ge T, M_\ell \subseteq F_T\}\hspace*{-1.1in}
         \quad
                & \hbox{because $|M_k|=k$}
                        \\
  &\;\leq\; \sum\{ |M_k| :  M_k\subseteq F_T\}
                &
                        \\
  &\;\leq\; |F_T|
                & \hbox{since the $M_k$'s are disjoint}\\
  &\;\leq\; sT.
                &\hbox{since $\barF$ is $s$-size-competitive}
\end{align*}
Thus, the bid set $\calB$ is $s$-competitive for universe $\calU=[m]$.

  \medskip

  We now present the construction of the metric space satisfying
  condition $(\ast)$. Let
  $\calC = [1]\times [2]\times ...\times [m]$, that is, $\calC$
 is the set of integer vectors
  $\baru = (u_1,u_2,\dots,u_m)$ where $ u_\ell \in [\ell ]$
         for all $\ell=1,2,\dots,m$.
  For each $\ell\in [m]$, introduce a set
  $M_\ell = \braced{f_{\ell,1},f_{\ell,2},\dots,f_{\ell,\ell}}$,
  and for each  node $\baru$ in $\calC$,
connect $\baru$ to $f_{\ell,u_\ell}$  with an edge of
  length $\delta_\ell = 1 + (m!)^{-\ell}$.
  The set of facilities is $\calF = \bigcup_{\ell=1}^m M_\ell$.
  All distances between points in $\calC\cup\calF$
  other than those specified above are determined
  by shortest-path lengths. Since $1<\delta_\ell\le 2$
  for all $\ell$, the resulting distance function 
  satisfies the triangle inequality.

  We have $\cost(M_j) =  m!\delta_j$ for each $j\in [m]$.
  We prove $(\ast)$ by contradiction. Fix some index $j$ and
  consider a set $F\subseteq\calF$ that does not contain
  $M_\ell$ for any $\ell\ge j$: for each $\ell\ge j$ there is $i_\ell\le \ell$
  such that $f_{\ell,i_\ell}\notin F$. Define a customer
  $\baru$ as follows: $u_i = 1$ for $\ell = 1,\dots,j-1$ and
  $u_i = i_\ell$ for $\ell = j,\dots,m$. Then the facility $f_{\ell,i}\in F$
  serving this $\baru$ must have  $\ell < j$ or $i\neq i_\ell$.
  Either way, it is at distance at least
  $\delta_{j-1}$ from $\baru$. Since every other customers pays
         strictly more than 1, we get
  $\cost(F) > m!-1 + \delta_{j-1}
  = m!\delta_j = \cost(M_j)$ -- a contradiction.
\end{proof}


\smallskip

\begin{proofof}{Theorems~\ref{thm: size competitive deterministic}(b)
    and Theorem~\ref{thm: size competitive randomized}(b)}
Lemma~\ref{lem: size lb} and the lower bounds for
online bidding in Theorem~\ref{thm: online bidding lower}
immediately imply Theorems~\ref{thm: size competitive deterministic}(b)
and~\ref{thm: size competitive randomized}(b).
\end{proofof}

Next we turn attention to polynomial-time algorithms. First,
we prove Theorem~\ref{thm: size competitive
  offline} --- that there exists a polynomial-time $O(\log
m)$-size-approximation algorithm for the offline problem. 
We will give a polynomial-time algorithm that, given $k$
and a problem instance, finds a
facility set of size $O(k\log m)$ and cost at most $\opt_k$.
(Here $m=|\calC|$ is the number of customers.)


\smallskip

\begin{proofof}{Theorem~\ref{thm: size competitive offline}}
Without loss of generality, we assume that each customer has distance
0 to its closest facility.  (Otherwise, for each customer $u$ we can 
subtract $d_{u\cal F}$ from the distance $d_{uf}$ to each facility.  This
decreases the cost of each facility set by a uniform amount, so any
solution having optimal cost in the modified instance will
also have optimal cost in the original instance.)

The algorithm first runs the bicriteria approximation algorithm from
\cite{young00kmedians} that, in time $O(k(m+n)\log(m/\epsilon))$,
finds a facility set of size
$O(k\log(m+m/\eps))$ and cost at most $(1+\eps)\opt_k$. This
algorithm is run with
$\eps=1/m$.  As a result, we obtain a facility set $F$ of size $O(k \log m)$
and cost at most $(1+1/m)\opt_k$.  We then greedily add to
$F$ a single facility $f$ that minimizes $\cost(F\cup\{f\})$.  The
algorithm returns $F\cup\{f\}$.

To finish, we show that the facility set $F\cup\{f\}$ has
size $O(k\log m)$ and cost at most $\opt_k$.
The size bound is immediate from the size bound on $F$.
To show the cost bound, note that if we add to $F$,
for the customer contributing the maximum amount to the current cost,
the closest facility $f$ to that customer, 
the cost for that customer would decrease to $0$.
(Recall our assumption that $d_{u\cal F}=0$ for each customer $u$.)  
Thus, adding this $f$ decreases the overall cost by at least
the current cost times $1/m$, and we get
$\cost(F\cup\{f\}) \le (1-1/m) \cost(F) \le (1-1/m)(1+1/m)\opt_k \le \opt_k$.
\end{proofof}


By the reduction in Lemma~\ref{lem: size reduction deterministic},
this gives a polynomial-time size-approximation algorithm
for the incremental problem:

\smallskip

\begin{proofof}{Theorem~\ref{thm: size competitive polynomial}}
By Theorem~\ref{thm: online bidding upper}(a),
there is a deterministic polynomial-time algorithm
for online bidding with competitive ratio $O(1)$.
Using this online bidding algorithm, and 
using Theorem~\ref{thm: size competitive offline}
to compute an $O(\log m)$-size-approximate $k$-median $F^*_k$
for each $k$, the reduction in Lemma~\ref{lem: size reduction deterministic}
gives an $O(\log m)$-size-competitive deterministic polynomial-time algorithm
for incremental medians.
\end{proofof}



\section{Incremental Cost-Competitive Medians}
\label{sec: cost-competitive medians}

In this section we prove
Theorems~\ref{thm: cost competitive deterministic}
and ~\ref{thm: cost competitive randomized}.

In the analysis we will use the following fact, whose proof
can be found in \cite{cky05} and is also implicit in \cite{jain01}.
(See also Sections~\ref{sec: lambda-relaxed metrics}
and~\ref{sec: fractional medians} for generalizations with proofs.)
Given two sets of facilities $A$ and $B$, 
let $\Gamma(A,B)$ be a set $C\subseteq B$ of cardinality
at most $|A|$ defined as follows. For each facility 
$g\in A$, choose a single customer $r(g)\in\calC$ to be defined shortly, and
let $f$ be the facility of $B$ serving $r(g)$. We define $C$
to be the set of such $f$'s as $g$ spans $A$, where $r(g)$ is chosen so as to
minimize $d_{gr(g)}+d_{r(g)f}$.

It will be convenient to introduce distances between facilities: given
two $f,g\in\calF$, let $d_{fg} = \min_{x\in\calC}(d_{fx}+d_{xg})$.
For $f\in \calF$ and $F\subseteq \calF$,
define the distance from $f$ to $F$ by
$d_{fF} = \min_{g\in F} d_{fg}$.
In other words, in order to define $C$, for each $a\in A$ we add to $C$ an
element of $B$ closest to $a$.


\begin{fact}\label{fact: cost of Fk}
Let $1\leq h < k\leq n$, and assume that the distance function is metric.
Consider an $h$-median $A$ and a $k$-median $B$,  and let $C=\Gamma(A,B)$. 
Then $|C|\leq h$, and $C$ is a subset of $B$ such that for every customer $u$ we have $d_{uC}\leq 2d_{uA}+d_{uB}$.
\end{fact}
This implies $\cost(C)\leq\cost(B)+2\cost(A)$.

To prove
Theorems~\ref{thm: cost competitive deterministic}
and ~\ref{thm: cost competitive randomized}, the argument is based on another reduction from incremental medians to online
bidding, presented in the lemma
below. This implies the theorems.


\begin{lemma}\label{lem: cost reduction}
Consider an instance of the metric medians problem.
Assume that for each $k$ we have a set of facilities
$F_k^*$ of size at most $k$ and cost at most $c\cdot\opt_k$.	
Let $\beta\geq 1$.
Assume that we have a (randomized)
polynomial-time $\beta$-competitive algorithm for online bidding.
Then in (randomized) polynomial time we can
compute an \emph{incremental} solution
$\barF = (F_1,...,F_n)$, where each $F_k$ has 
size at most $k$ and (expected) cost at most $2\beta c\cdot\opt_k$.
\end{lemma}

\begin{proof} 
Without loss of generality
$\cost(F^*_k) \ge \cost(F^*_{k+1})$ for all $k<n$.
  The algorithm constructs the incremental solution
 $\barF$ from $F_1^*,...,F_n^\ast$ 
 in several steps.  First, fix some indices $1 = \kappa(1) < \kappa(2) < \ldots < \kappa(m)$
by a method to be described later, and let
$\calK\subseteq[n]$ denote the set of indices.

  Next, compute the sets $F_{\kappa (i)}$ as follows.
  $F_{\kappa(m)} = F^*_{\kappa(m)}$.
  For $i=m-1,m-2,...,1$, inductively
  define $F_{\kappa(i)}$ to contain
  the facilities within $F_{\kappa(i+1)}$
  that are ``closest'' to $F^*_{\kappa(i)}$ in the following sense.
With the $\Gamma(\cdot,\cdot)$ notation seen above, we define
$F_{\kappa(i)}=\Gamma(F^*_{\kappa(i)},F_{\kappa(i+1)})$.
Directly from the definition,
$|F_{\kappa(i)}|\le|F^*_{\kappa(i)}|\leq \kappa (i)$.
  
Finally, for indices $k\notin\calK$, define $F_k = F_{\kappa (i)}$,
where $i$ is maximum such that $\kappa (i)\leq k$
(this $i$ is well defined, since $\kappa(1) = 1$.)
Obviously, $|F_k|\leq \kappa(i)\leq k$.
To complete the construction, it remains to describe how
to compute $\calK$, which we momentarily defer.

\smallskip

\noindent   
To analyze the cost, for a given $k$,
 let $i$ be maximum such that $\kappa (i)\leq k$. From 
the definition of $F_{\kappa (i)}$ and Fact~\ref{fact: cost of Fk} summed
over all $u\in \calC$, we have:
$\cost (F_{\kappa(i)})
         \leq 2\cost (F^*_{\kappa(i)})
                        +\cost (F_{\kappa(i+1)})$.
Applying this inequality repeatedly gives:
\begin{eqnarray*}
\cost(F_k) \;=\; \cost(F_{\kappa (i)})
        \;\le\; 2\sum_{j=i}^m \cost(F^*_{\kappa (j)}).
\end{eqnarray*}
To complete the proof of the lemma,
it suffices to define $\calK$ so that
\begin{equation}\label{eqn: cost}
    \sum_{j=i}^m \cost(F^*_{\kappa(j)})
        \;\le\; \beta\,\cost(F^*_k).
\end{equation}
since this will imply
$\cost(F_k) \le 2\beta \cost(F^\ast_k) \le 2\beta c\cdot \opt_k$.

We now prove (\ref{eqn: cost}).
Let $\calU=\{\cost (F^*_{k}): 1\leq k\leq n \}$ and
take $\calB$ to be the set of bids used by the $\beta$-competitive 
online bidding algorithm  for universe $\calU$. We define
$\calK=\{ \kappa: \cost(F^*_\kappa) \in \calB \}$,
with ties broken in favor of smaller indices (ties may happen if several
sets of facilities have the same cost). Note that with this
tie-breaking rule, we always have $1\in {\calK}$.
(If the online bidding algorithm
is randomized, then $\calB$ and therefore $\calK$ are random.)
Then the left-hand side of~(\ref{eqn: cost}) is exactly the sum of
the bids paid by the online bidding algorithm for
threshold $T=\cost(F^*_k)$.  Since $\calB$ is
$\beta$-competitive, this cost is at most $\beta\,\cost(F^*_k)$,
so~(\ref{eqn: cost}) holds.
This completes the proof.
\end{proof}


The algorithm of Theorem~\ref{thm: cost competitive deterministic}(a) is
obtained by combining the reduction
used in the proof of Lemma~\ref{lem: cost reduction} with the
deterministic $4$-competitive online bidding algorithm given in the proof
of Theorem~\ref{thm: online bidding upper}(a).
The resulting algorithm is detailed in the displayed
Algorithm~\ref{alg:cost-competitive}.

\begin{algorithm}[ht]
        \caption{The 8-cost-competitive deterministic incremental 
                        median algorithm.}
        \label{alg:cost-competitive}
\begin{algorithmic}
 \FORALL{  $k =1,...,n$,}
  \STATE compute a $k$-median solution $F^*_k$
  \ENDFOR
 \STATE $\calK\leftarrow \{ 1\}$
 \FORALL{$k =2,...,n$,}
    \STATE  if $\cost (F^*_k)=0< \cost (F^*_{k-1})$ then add $k$ to $\calK$
    \STATE if  $\cost(F^*_k)>0$ then if 
$\lceil \log \cost (F^*_k)\rceil < 
 \lceil \log \cost (F^*_{k-1})\rceil$ then add $k$ to $\calK$
 \ENDFOR
 \STATE $\ell\leftarrow \max (\calK )$; $F_\ell \leftarrow F^*_\ell$
 \STATE{ for all $k = \ell+1,...,n$ do $F_{k} \leftarrow F_\ell$}
 \FOR{ each $k\in \calK-\braced{\max (\calK)}$, in decreasing order}
   \STATE $F_k\leftarrow \Gamma (F^*_k,F_\ell)$
   \STATE { for all $k' = k+1,...,\ell-1$ do $F_{k'} \leftarrow F_k$ }
   \STATE $\ell \leftarrow k$
   \ENDFOR
\end{algorithmic}
\end{algorithm}


\section{$\lambda$-Relaxed Metrics}
\label{sec: lambda-relaxed metrics}


In this section we prove
Theorem~\ref{thm: cost competitive lambda deterministic},
namely the upper bounds for cost-competitive ratios
when the distance function is a $\lambda$-relaxed metric.
For $\lambda\ge 1$,
by a $\lambda$-relaxed metric, we mean a distance
function that satisfies
$d_{yf}\le \lambda (d_{xf} + d_{xg} + d_{yg})$
for all $f,g\in \calF$ and $x,y\in \calC$.

We start with a generalization of Fact~\ref{fact: cost of Fk}
from the previous section. Recall that
$\Gamma(A,B)$ is the set of up to $|A|$ elements of $B$
that are closest to the elements of $A$
(see the definition in previous section.)


\begin{lemma}\label{fact: cost of Fk: relaxed}
Assume that the distance function is a $\lambda$-relaxed metric.
Given two sets of facilities $A$ and $B$, let $C=\Gamma(A,B)$.
Then for every customer $u\in\calC$ we have
$d_{uC}\leq 2\lambda d_{uA}+\lambda d_{uB}$.
\end{lemma}

\begin{figure}[t]
\begin{center}
\includegraphics[width=1.9in]{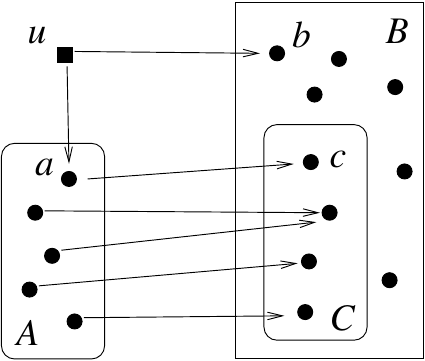} %
\caption{Illustration of Lemma~\ref{fact: cost of Fk: relaxed}}.
\label{fig: lambda-relaxed}
\end{center}
\end{figure}

\begin{proof}
For a given $u\in \calC$, choose
$a\in A$ such that $d_{uA}=d_{ua}$,
$b\in B$ such that $d_{uB} = d_{ub}$, and
$c \in C$ such that $d_{aB}=d_{ac}$,
(See Figure~\ref{fig: lambda-relaxed}.)
By the definition of the $\lambda$-relaxed metric,
for every customer $x\in\calC$ we have 
$d_{uc}\leq \lambda (d_{ua}+d_{xa}+d_{xc})$.
There is $x\in\calC$ for which $d_{ac} = d_{xa}+d_{xc}$,
and choosing this $x$ we get
$d_{uc}\leq \lambda (d_{ua}+d_{ac})$.
Thus:
\begin{equation*}
d_{uC} \;\leq\; d_{uc}
         \;\leq\; \lambda (d_{ua}+d_{ac})
         \;\leq\; \lambda (d_{ua}+d_{ab})
         \;\leq \; \lambda (d_{ua}+ (d_{ua}+d_{ub}))
        \;=\; 2\lambda d_{uA}+\lambda d_{uB},
\end{equation*}
completing the proof.
\end{proof}

\begin{proofof}{Theorem~\ref{thm: cost competitive lambda deterministic}}
The algorithm used to prove Theorem~\ref{thm: cost competitive lambda deterministic} 
is the same as in  the metric case (see the proof of Lemma~\ref{lem: cost reduction}), 
except for the definition of ${\cal B}$.
Let ${\cal U}=\{ \cost (F^*_{k}): 1\leq k\leq n \}$.

${\cal B}$ contains 0 iff ${\cal U}$ does, plus the following
elements: 
in the deterministic case, for every $i\in\integers$,
${\cal B}$ contains the
maximum element in the set ${ \cal U}\cap [0, (2\lambda)^i]$
(if it is non-empty); in the randomized case, pick a real number
$\xi\in (0,1]$ uniformly at random, and for every $i\in\integers$,
${\cal B}$ contains the
maximum element in the set ${ \cal U}\cap  [0,e^{\xi} (e\lambda)^i]$ 
(if it is non-empty).

The analysis is similar to the proof of 
Theorem~\ref{thm: cost competitive deterministic}.
Let $\calK=\{\kappa(i): 1\leq i\leq  |\calB | \}$  be 
defined from ${\calU}$ and ${\calB}$ as in the
proof of Theorem~\ref{thm: cost competitive deterministic}.
Choose $i$ to be the maximum index such that $\kappa(i)\leq k$.
From Lemma~\ref{fact: cost of Fk: relaxed} summed over
$u\in\calC$, we have:
$\cost (F_{\kappa(i)})
        \leq 2\lambda \cost (F^*_{\kappa(i)})
                +\lambda\cost (F_{\kappa (i+1)})$.
Applying this inequality repeatedly gives
\begin{equation*}
\cost (F_{k}) \;=\; \cost (F_{\kappa(i)})
         \;\leq\; 2\lambda
          \sum_{j= i}^m \lambda^{j-i}\cost (F^*_{\kappa(j)}). 
\end{equation*}
To continue the analysis in the deterministic case, let $p$ be such that
$(2\lambda)^{p-1} < \cost (F^*_{\kappa(i)}) \le (2\lambda)^{p}$.
By the definition of $\kappa(i)$, we have:
\begin{eqnarray*}
\cost (F_{k})
         \;\leq \; 2\lambda \sum_{j\geq i} \lambda^{j-i}
                        (2\lambda)^{p-j+i}
        \;=\; 2 (2\lambda )  (2\lambda)^{p}
        \;\leq\; 8 \lambda^2 \cost (F^*_{k}),
\end{eqnarray*}
since $k<\kappa (i+1)$ and so $\cost (F^*_{k})\geq  (2\lambda)^{p-1}$.
Hence our algorithm is $8\lambda^2$-cost-competitive
when $F^*_{k}$ is the optimal $k$-median and is
$8\lambda^2c$-cost-competitive when $F^*_{k}$ is a $c$-approximation.

In the randomized case, let $p$ such that
$e^{\xi } (e\lambda)^{p-1}
         < \cost (F^*_{\kappa(i)})
         \le e^\xi (e\lambda)^{p}$.
By the definition of $\kappa(i)$, we have:
\begin{equation*}
\cost (F_{k})
        \;\leq\; 
        2\lambda   e^\xi\sum_{j\geq i} \lambda^{j-i}
        (e\lambda)^{p-j+i}
        \;=\; 
        \frac{1}{1-1/e} 2\lambda   e^\xi  (e\lambda)^{p}.
\end{equation*}
As in the proof of
Theorem~\ref{thm: online bidding upper}(b), the
expected value of $ e^\xi  (e\lambda)^{p}/ \cost (F^*_{k})$ is
distributed like $e^\xi\lambda$ where $\xi$ is uniform in $(0,1]$, so
\begin{equation*}
E[\cost (F_{k})]
         \;\leq\; \frac{1}{1-1/e} (2\lambda )\cost (F^*_{k}) \lambda(e-1)
        \;=\;      2e \lambda^2\cost (F^*_{k}),
\end{equation*}
and the theorem follows.
\end{proofof}


\section{Incremental Fractional Medians}
\label{sec: fractional medians}

In this section we prove Theorem~\ref{thm: fractional incremental median}.

A fractional $k$-median is given by a feasible solution to the
following linear program relaxation of the
standard $k$-median integer linear program:
\[
\mbox{minimize}_{x,y}~ \sum_{f\in\calF,u\in\calC} d_{uf}x_{uf} \mbox{~~subject to~}
  \left\{\begin{array}{rcll}
    \displaystyle
      \sum_{f\in\calF} x_{uf} & = & 1
    & \forall u\in\calC
    \\
      \displaystyle
      x_{uf} & \le & y_f
    & \forall u\in\calC,f\in\calF
    \\
    \displaystyle
    \sum_{f\in\calF} y_f & \le & k
    \\
    x_{uf} & \ge & 0  
    & \forall u\in\calC,f\in\calF.
  \end{array}\right.
\]
In this section we let $x$ denote a fractional median.
The \emph{capacity required by $x$ at $f$} is
defined as $|x|_f = \max_{u\in\calC}x_{uf}$
(essentially, the value of $y_f$ in the linear program).
The total \emph{capacity required by $x$} is
$|x| = \sum_{f\in\calF} |x|_f$, and
the \emph{cost of $x$} is the objective function value
$\cost(x)=\sum_{f\in\calF,u\in\calC} d_{uf}x_{uf}$.
Recall that for two fractional medians $x$, $x'$, we say
that $x$ \emph{dominates} $x'$ if
$|x'|_f \le |x|_f$ for all $f\in \calF$.


\begin{lemma}\label{fact: cost of fractional}
Let $1\le h\leq k\le n$, and
assume that the distance function is metric.
Consider two fractional medians, a fractional $k$-median $x$
and a fractional $h$-median $z$.
There exists a fractional $h$-median $x'$ dominated by $x$
such that $\cost(x')\le \cost(x)+2\cdot\cost(z)$.
\end{lemma}

\begin{proof}
Construct $x'$ as follows.
For each facility $g\in\calF$, choose
(by a method to be described later)
 a single customer $\randcust{g}\in\calC$ 
``responsible'' for $g$.
For each $u\in\calC$ and $f\in\calF$,
take $x'_{uf} = \sum_{g\in\calF} z_{ug}x_{\randcust{g}f}$.

\smallskip
For each $u\in\calC$, applying the first constraint
of the above linear program, we have
$\sum_{f\in\calF} x'_{uf} 
        = \sum_{f,g\in\calF} z_{ug}x_{\randcust{g}f}
        = \sum_{g\in\calF}(\sum_{f\in\calF} x_{\randcust{g}f}) z_{ug}
        = \sum_{g\in\calF}z_{ug} = 1$,
so $x'$ is a valid fractional median (for some capacity value).

\smallskip
Next we prove that $|x'|$ is a fractional $h$-median. The intuition is that
``routing'' $x'$ through $z$ (as described later) ensures this.
To prove that $x'$ is a $h$-median, we bound the total capacity required by $x'$:
\begin{eqnarray*}
|x'| &=&
\sum_{f\in\calF} \max_{u\in\calC} \sum_{g\in\calF} z_{ug} x_{\randcust{g} f}
                        \\
&\le&
\sum_{g,f\in\calF} \max_{u\in\calC} z_{ug} x_{\randcust{g} f}
                        \\
&=&
\sum_{g\in\calF} \Big(\sum_{f\in\calF} x_{\randcust{g} f}\Big) \max_{u\in\calC} z_{ug} 
                        \\
&=&
\sum_{g\in\calF} \max_{u\in\calC} z_{ug}
           \;=\; |z| \;\le\; h.
\end{eqnarray*}

\smallskip
We now claim that
the fractional median $x'$ is dominated by $x$. Indeed,
fixing an $f\in\calF$, for any $u\in\calC$ we have 
$x'_{uf}  = \sum_{g\in\calF} z_{ug} x_{\randcust{g}f}
        \le \sum_{g\in\calF} z_{ug} \cdot\max_{w\in\calC} x_{wf}
         \le \max_{w\in\calC} x_{wf} = |x|_f$;
therefore $|x'|_f \le |x|_f$, as claimed.

\smallskip
To finish the proof, we show that we can choose the responsible customers so that
$\cost(x') \le 2\cost(z)+\cost(x)$.
Consider choosing $\randcust{g}$ randomly for each $g\in\calF$ so that
$\Pr[\randcust{g}=w] = z_{wg}/\sum_{u\in\calC}z_{ug}$
(if the denominator is zero, choose $\randcust{g}$ arbitrarily).
Now bound the expected cost of $x'$ as follows.
Imagine ``routing'' one unit of weight from each customer $u\in\calC$ 
to the facilities in stages:
(stage 1) send $z_{ug}$ units from each $u\in\calC$ to each $g\in\calF$;
(stage 2) from each $g\in\calF$, send all arriving weight to $\randcust{g}\in\calC$;
(stage 3) from each $w\in\calC$, split all arriving weight and send an $x_{wf}$ 
fraction to each $f\in\calF$.

For each $u\in\calU$ and $f\in\calF$,
an easy calculation shows that for every choice of $r(g)$, $x'_{uf}$ units of the weight 
that leaves $u$ at the start end up at $f$ at the end.

Since $d_{uf} \le d_{ug}+d_{\randcust{g}g}+d_{\randcust{g}f}$, the cost of $x'$ is at 
most the sum of the costs of the stages, where the cost of sending weight between 
any $u\in\calC$ and any $f\in\calF$ in a stage is $d_{uf}$ per unit.

The first stage costs $\sum_{u\in\calC,g\in\calF} d_{ug}z_{ug} = \cost(z)$.

In the second stage, for each $g\in\calF$, the total weight to be sent is $\sum_{u\in\calC}z_{ug}$.
Therefore, with the random choice of $\randcust{g}$,
the expected weight sent to any given $w\in\calC$ is 
$(\sum_{u\in\calC}z_{ug})\Pr[\randcust{g} = w] = z_{wg}$
(using the definition of $\randcust{g}$).
Thus, the expected cost of the second stage is
$\sum_{g\in\calF,w\in\calC}d_{wg}z_{wg} = \cost(z)$.

In the third stage, for each $w\in\calC$, the expected weight to be split and 
sent on is $\sum_{g\in\calF} z_{wg}=1$
(using from above that the expected weight sent from $g\in\calC$ to $w$ in the second 
stage is $z_{wg}$).
The fraction of this weight sent to each $f\in\calF$ is $x_{wf}$,
so the expected weight sent from $w$ to $f$ is $x_{wf}$.
Thus, the expected cost of the third stage is
$\sum_{w\in\calC,f\in\calF}d_{wf}x_{wf} = \cost(x)$.

\smallskip

In sum, the expected total cost of the stages is at most $2\cost(z)+\cost(x)$.
Since the cost of the stages is an upper bound on the expectation of $\cost(x')$,
we conclude $E[\cost(x')] \le 2\cost(z)+\cost(x)$.
So there is some way to choose the responsible customers so that 
$\cost(x') \le 2\cost(z)+\cost(x)$.
\end{proof}


\medskip

\begin{proofof}{Theorem~\ref{thm: fractional incremental median}}
With the lemma in place, the proof of the theorem is
essentially the same as the proof of Theorems~\ref{thm: cost competitive deterministic}
and ~\ref{thm: cost competitive randomized},
along with a minor observation
about fractional strategies being closed under randomization.
More precisely, recall that a $c$-cost-competitive incremental fractional median
gives, for every integer $k\in[n]$, a fractional
$k$-median $x^{k}$ with $\cost(x^k) \le c\cdot\fopt_k$,
such that $x^{k}$ is dominated by $x^{k+1}$ for all $k<n$.  

We first show that a $c$-cost-competitive fractional
median with the minimum ratio $c$ can be computed in polynomial
time using linear programming as follows.
For each $k\in[n]$, compute $\fopt_k$ (using
linear programming). Then,
set up a linear program that has the following
constraints: (i) the fractional $k$-median
constraints for each $x^k$,
(ii) the incrementality constraints saying that
each $x^{k+1}$ dominates $x^k$ for $k<n$, and
(iii) the cost-competitiveness constraints saying that
$\cost(x^k) \le c\cdot \fopt_k$, for each $k$.
The objective function is to minimize $c$.
The solution of this linear program is a
$c$-cost-competitive incremental fractional median with minimum $c$.

It remains to show that $c\le 2e$, by proving that there exists a
$2e$-cost-competitive fractional incremental median.
 We first show
that $c\le 8$, using the existence of a deterministic
$4$-competitive online bidding algorithm.

First, fix some indices $1 = \kappa(1) < \kappa(2) < \ldots < \kappa(m)$
by a method to be described later, and let
$\calK\subseteq[n]$ denote this set of indices.

Next, compute the fractional medians $x^k$ as follows.
Let $z^k$ denote an optimal fractional $k$-median.
We initialize $x^{\kappa(m)} = z^{\kappa(m)}$.
Then, for
$i = m-1, m-2, ..., 1$, inductively define $x^{\kappa(i)}$
to be a minimum-cost fractional $\kappa(i)$-median
among those dominated by $x^{\kappa(i+1)}$.
Finally, for indices $k\notin\calK$,
define $x^{k} = x^{\kappa(i)}$, where $i$
is maximum such that $\kappa(i) \le k$.
To complete the construction, it remains to describe how to compute $\calK$,
which we momentarily defer.

To analyze the required capacity, note that
for $ \kappa(i) \le k < \kappa(i+1)$ we have
$|x^k| = |x^{\kappa(i)}| \le \kappa(i) \le k$,
and thus $x^k$ is indeed a fractional $k$-median.

To analyze the cost, for a given $k$, let $i$ be
maximum such that $\kappa(i)\leq k$. From 
Lemma~\ref{fact: cost of fractional} we have
$\cost(x^{\kappa(i)})\le \cost(x^{\kappa(i+1)})+2\cdot \fopt_{\kappa(i)}$.
Applying this inequality repeatedly gives:
\begin{equation}
 \cost(x^{k})=\cost(x^{\kappa(i)})~\le~2\sum_{j=i}^m \fopt_{\kappa(j)}.
                \label{eqn: cost of fractional}
\end{equation}
To complete the proof, it suffices to define $\calK$ so that
\begin{equation}\label{eqn: cost2}
    \sum_{j=i}^m \cost(x^{\kappa(j)})
        \;\le\; \beta\,\cost(z^k)
\end{equation}
since this will imply
$\cost(x^k) \le 2\beta \cost(z^k) \le 2\beta c\cdot \opt_k$.

We will now prove~(\ref{eqn: cost2}) for $\beta=4$.
Let $\calU=\{\fopt_k: 1\leq k\leq n \}$ and
take $\calB$ to be the set of bids used by the $4$-competitive 
online bidding algorithm  for universe $\calU$.
Let $\calK=\{\kappa(i): 1\leq i\leq  m \}$  be a minimal
set (containing 1) such that
$\calB = \{\fopt_{\kappa(i)}: 1\leq i\leq  m\}$ (breaking
ties in favor of smaller indices). Then the left hand side of~(\ref{eqn: cost2}) is exactly the sum of
the bids paid by the online bidding algorithm for threshold $T=\cost(z^k)$. Since $\calB$ is
$\beta$-competitive, this cost is at most $\beta\cost(z^k)$, so~(\ref{eqn: cost2}) holds. This completes the
proof of 8-competitiveness.

To improve this ratio to $2e$, consider carrying out the above construction 
using the $e$-competitive randomized online bidding algorithm.
That algorithm generates a random bidding set $\calB$ for our universe $\calU$.
Applying the construction above gives us a random incremental fractional
solution $(x^{k})_k$.
For each $k$, $f$ and $u$, take $\barx^{k}_{uf}$ to be the expectation
of $x^{k}_{uf}$ for this random solution.
This gives an incremental fractional solution
which, for each $k$, has $|\barx|\le k$,
and whose cost is equal to the expected cost of the random solution.
Thus, $(\barx^{k})_k$ is $2e$-cost-competitive.

Summarizing, we showed that
there exists a $2e$-cost-competitive fractional incremental median.
As the algorithm given earlier computes a $c$-cost-competitive fractional
incremental median that minimizes $c$, 
the theorem follows.
\end{proofof}


\section{Incremental Algorithms for $kl$-Medians}
\label{sec: kl-medians}

In this section we prove Theorem~\ref{thm: kl-medians deterministic}.
Recall that in the $kl$-median problem, for given
$1\le k < l \le n$, we need to compute
two facility sets $F_k \subseteq F_l$ with $|F_k| = k$
and $|F_l| = l$, minimizing the cost-competitive ratio
\begin{eqnarray*}
c &=& \max\braced{ \frac{\cost(F_k)}{opt_k} ,
                        \frac{\cost(F_l)}{opt_l } }.
\end{eqnarray*}
We now prove that the optimal ratio $c$
for this problem is between $2-1/(l-k+1)$ and $2-1/l$. 


\begin{proofof}{Theorem~\ref{thm: kl-medians deterministic}(a)}
We start with the upper bound proof. Our method
here is very different from the previous bounds in this
paper and it does not rely on online bidding. 
Let $F$ and $G$ denote, respectively, the optimum $k$-median and 
the optimum $l$-median.
Without loss of generality, we can assume that
$F\cap G = \emptyset$, for otherwise we can duplicate
the facilities in $F\cap G$. Our algorithm 
chooses the better of two options below
(the one with the better ratio):
\begin{description}
\item{(i)} $F_k = F$, and $F_l$ is a set with minimum $\cost(F_l)$, 
such that $|F_l|=l$ and $F\subseteq F_l\subseteq \calF$,
or
\item{(ii)} $F_k$ is a set with minimum $\cost(F_k)$
such that $|F_k|=k$ and $F_k\subseteq G$,
and $F_l = G$.
\end{description}

We now show that this algorithm's competitive ratio is
at most $2-1/l$. It is sufficient to show 
that there exists a $k$-element set $X\subseteq G$ such that

\begin{eqnarray}
\cost(X) + \cost(F\cup G-X) &\le& (2-1/l)[\cost(F)+\cost(G)].
                \label{eqn: kl-medians claim}
\end{eqnarray}

Indeed, Inequality (\ref{eqn: kl-medians claim})
implies that at least one of the following
two options must hold:
either $\cost(X) \le (2-1/l) \cost(F)$, or
$\cost(F\cup G-X) \le (2-1/l)\cost(G)$. 
In the first case our algorithm can choose option (ii),
and then we will have $\cost(F_k) \le \cost(X) \le (2-1/l) \cost(F)$.
In the second case it can choose option (i),
and then $\cost(F_l) \le \cost(F\cup G-X) \le (2-1/l)\cost(G)$. 
In both cases, the algorithm's ratio is at most $2-1/l$.

It remains to show that there is a $k$-element
set $X\subseteq G$ that satisfies (\ref{eqn: kl-medians claim}).
Our proof is based on a probabilistic argument.
We start with some notation.
For each $f\in F$ and $g\in G$, denote by $C_{fg}$ the
set of customers that are served by $f$ when $F$ is the
facility set and by $g$ when $G$ is the facility set.
Let $\weight_{fg}$ be the cardinality of $C_{fg}$.
By $a_{fg}$ (resp. $b_{fg})$ we denote the average distance between
a customer $x\in C_{fg}$ and $f$ (resp. $g$.)
Formally, $a_{fg} = \sum_{u\in C_{fg}} d_{uf}/|C_{fg}|$ and
$b_{fg} = \sum_{u\in C_{fg}} d_{ug}/|C_{fg}|$.
It is convenient to think of
$C_{fg}$ as a single point with weight $\weight_{fg}$ whose distances
to $f$ and $g$ are $a_{fg}$ and $b_{fg}$, respectively.
The costs of $F$ and $G$ can then be written as
$\cost(F) = \sum_{f\in F,g\in G}\weight_{fg}a_{fg}$,
and $\cost(G) = \sum_{f\in F,g\in G}\weight_{fg}b_{fg}$.

We now define a probability distribution on $k$-element
subsets of $G$. Let $\weight_{fG} = \sum_{g\in G} \weight_{fg}$.
For each $f\in F$ and $g\in G$, define
$\barweight_{fg} = \weight_{fg}(\weight_{fG}-\weight_{fg})^{-1}$.
(Assume for now that $\weight_{fg}>0$ for all $f,g$.
We will explain later how to extend the argument to the
general case.) Choose a random mapping
$\pi: F \to G$ as follows: for any $f\in F$ set $\pi(f) = g$
with probability
$\barweight_{fg}/\sum_{h\in G}\barweight_{fh}$.
For any such mapping $\pi$ let $X_\pi$ be the $k$-element subset of
$G$ that consists of $\pi(F)$ and arbitrary
$k-|\pi(F)|$ elements of $G-\pi(F)$.
(Intuitively, we would like to take our random $k$-set $X\subseteq G$
to be $\pi(F)$, but then $F\cup G - \pi(F)$ may have cardinality larger
than $l$ and will not be a valid $l$-median. This is why
we add these additional elements to $X_\pi$.)

Consider some $f\in F$ and $h\in G$. The cost of
of serving $C_{fh}$ from $F\cup G-X_\pi$ is at most $w_{fh} a_{fh}$.
If $h\notin X$, we can also bound this cost by $w_{fh} b_{fh}$.
We also want to estimate the cost of serving $C_{fh}$ from $X_\pi$.
If $h\in X_\pi$ then this cost is at most $w_{fh} b_{fh}$.
If $h\in G - X_\pi$ then, for any $g\in X_\pi$,
the cost of serving $C_{fh}$ from $X_\pi$ is bounded by the
cost of serving $C_{fg}$ from $g$, and thus (by the triangle inequality) it is at most
$w_{fh}a_{fh} + w_{fh}\min_{u\in C_{fg}}(d_{uf}+d_{ug})
        \le w_{fh}( a_{fh}+ a_{fg} + b_{fg})$.
Using $g = \pi(f)$ and summing over all $f\in F$ and $h\in G$, we get
\begin{eqnarray*}
\cost(X_\pi) + \cost(F\cup G-X_\pi) &\le&
        \sum_{f\in F}\Big[
               \sum_{h\in G-X_\pi} \weight_{fh}( a_{fh}+ a_{f\pi(f)}
                                        + b_{f\pi(f)}+b_{fh})
                                \\
        && \mbox{\hspace{1.75in}}
                +\;
                \sum_{h\in X_\pi} \weight_{fh} (a_{fh} + b_{fh})
                        \Big]
                        \\
        &=&
        \cost(F) + \cost(G)
          + \sum_{f\in F}\sum_{h\in G-X_\pi}
                \weight_{fh}(a_{f\pi(f)} + b_{f\pi(f)})
                        \\
        &\le&
        \cost(F) + \cost(G)
          + \sum_{f\in F}(\weight_{fG} - \weight_{f\pi(f)})
                        (a_{f\pi(f)} + b_{f\pi(f)}).
\end{eqnarray*}
Now, by the linearity of expectation, we have
\begin{eqnarray*}
E[\,\cost(X_\pi) + \cost(F\cup G-X_\pi)\,]
         &\le& \cost(F) + \cost(G)
                \\
         && +\;\sum_{f\in F}\sum_{g\in G}
        \frac{\barweight_{fg}}{\sum_{h\in G}\barweight_{fh}}
                 (\weight_{fG}-\weight_{fg})(a_{fg}+b_{fg})
                \\
         &=& \cost(F) + \cost(G)
                + 
                \sum_{f\in F}
                \frac{1}{\sum_{h\in G}{\barweight_{fh}}}
                         \sum_{g\in G}\weight_{fg} (a_{fg}+b_{fg})
                \\
        &\le& (2-1/l)[\cost(F) + \cost(G)].
\end{eqnarray*}
The last inequality holds because, for each $f\in F$,
${\sum_{h\in G}\barweight_{fh}}$
is minimized when $\weight_{fh} = \weight_{fG}/l$ for
all $h\in G$, and thus
$(\sum_{h\in G}\barweight_{fh})^{-1} \le 1-1/l$.
We conclude that there exists a set
$X_\pi$ that satisfies $(\ast)$, as claimed.

To complete the proof, we still need to extend the
argument to the general case, when some weights $w_{fg}$
are zero. Suppose first that we allow arbitrary positive
weights. Choose an arbitrarily small
$\epsilon > 0$ and set all zero weights $w_{fg}$ to
$\epsilon$ instead. The earlier argument implies that
for each $\epsilon$ there is $X$ that
satisfies $(\ast)$. Since there are finitely
many choices for $X$, there is $X$ that satisfies
$(\ast)$ for infinitely many values of $\epsilon$.
The continuity of $\cost(X) + \cost(F\cup G-X)$
with respect to $\epsilon$ implies that
$(\ast)$ holds for $\epsilon = 0$ as well.

In our case the weights are integer, so we cannot
use arbitrarily small weights. Instead, we create
a large number of copies of each customer (same for each),
and then add one customer to each empty set $C_{fg}$.
Then the same asymptotic argument as above applies.
\end{proofof}


\begin{proofof}{Theorem~\ref{thm: kl-medians deterministic}(b)}
Our lower bound is
a slight refinement of the one in \cite{MetPla00,mettu03}.
Let $k=1$ and $1 < l \le n$.
Consider the metric space $M$ with customers
$v_1,...,v_l$ and facilities $f_1,g_1,g_2,...,g_l$.
Each customer $v_j$ is
connected to facility $g_j$ by an edge of length
$\delta =1/l$. All customers are also connected to
facility $f_1$ with edges of length $1$.
(See Figure~\ref{fig: kl-medians}.) All other distances
are measured along the above defined edges.

\begin{figure}[h]
\begin{center}
\includegraphics[width=2.5in]{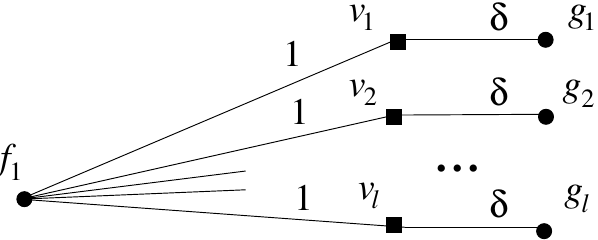} %
\caption{The metric space in the lower bound.
Facilities are represented by squares and customers
by circles.}\label{fig: kl-medians}
\end{center}
\end{figure}

Let $G = \braced{g_1,\dots,g_l}$. We have
$\cost(f_1) =l$ and $\cost(G) = l\delta$. Further,
for each $i$, we have
$\cost(g_i) = \delta + (l-1)(2+\delta)$ and
$\cost(G-\braced{g_i}\cup \braced{f_1}) = (l-1)\delta + 1$.
Thus, substituting $\delta = 1/l$, we get
$\cost(g_i)/\cost(f_1) = 2-1/l$ and
$\cost(G-\braced{g_i}\cup \braced{f_1})/\cost(G) = 2-1/l$
for all $i$.
So, if an incremental algorithm chooses $F_1 = \braced{g_i}$,
for some $i$, the ratio is at least $2-1/l$.
On the other hand, if it chooses $F_1 = \braced{f_1}$,
then $F_l = G-\braced{g_i}\cup\braced{f_1}$, for some $i$, and the ratio
is again at least $2-1/l$.

This completes the lower bound proof for $k=1$.
For $1< k < l$, we use the above construction with
$l' = l-k+1$ instead of $l$. We add 
$k-1$ facilities $f_2,...,f_k$ at a very
large distance from the above space and each other, and $k-1$
customers $v_{l+1} = f_2, ... ,v_{l+k-1}=f_k$.
Any $k$-median must include $f_2,...,f_k$, and
thus the argument above applies to this new space.
\end{proofof}


\section*{Acknowledgments}
We are grateful to anonymous referees for suggestions
to improve the presentation. We also wish to thank Yossi Azar
for pointing out references to previous work on online bidding
and simplifying the proof of
Theorem~\ref{thm: online bidding lower}(a).


\end{document}